\documentclass[fleqn,twoside]{article}

\usepackage[headings]{espcrc2}

\usepackage{amsmath}
\usepackage[dvips]{graphicx}
\usepackage{epsfig}

\title{Hyperon Physics from Lattice QCD}

\author{Huey-Wen Lin\address[JLab]{Thomas Jefferson National Accelerator Facility, Newport News, VA 23606}
        \thanks{Speaker}
 }

\runtitle{Hyperon Physics from Lattice QCD}
\runauthor{H.-W. Lin}

\begin{document}

\begin{abstract}
I review recent lattice calculations of hyperon physics, including hyperon spectroscopy, axial coupling constants, form factors and semileptonic decays.
\vspace{1pc}
\end{abstract}

\maketitle

\section{Introduction}

The hyperons are extremely interesting because they provide an ideal system in which to study SU(3) flavor symmetry breaking by replacement of up or down quarks in nucleons by strange ones. Hyperon semileptonic decays provide an additional way of extracting the CKM matrix element $V_{us}$ and offer unique opportunities to understand baryon structure and decay mechanisms. However, since hyperons decay in less than a nanosecond under weak interactions, their experimental study is not as easy as for the nucleons, and thus hyperon properties are not as well determined. Recent successes of lattice QCD in computing nucleon structure\cite{Orginos:2006pl,Hagler:2007hu,Zanotti:2008} provide assurance that lattice QCD can reliably predict the properties of hyperons as well. Knowledge of these properties can be very valuable in understanding hypernuclear physics, the physics of neutron stars and the structure of nucleons.

In this proceeding, I review the latest progress on the hyperon spectroscopy, axial coupling constants, electric charge radii, magnetic moments and semileptonic decay form factors using a mixed action, in which the sea and valence fermions use different lattice actions. Lattice calculations of hyperon scattering lengths are not included in this proceeding; for more details, please see Ref.~\cite{Beane:2006gf}. In our case, the sea fermions are 2+1 flavors of staggered fermions (in configuration ensembles generated by the MILC collaboration\cite{Bernard:2001av}), and the valence fermions are domain-wall fermions (DWF). The pion mass ranges from 300 to 700~MeV in a lattice box of size 2.6~fm. The gauge fields are hypercubic-smeared and the source field is Gaussian-smeared to improve the signal. Details on the configurations can be found in Ref.~\cite{WalkerLoud:2008bp}. For  three-point functions, the source-sink separation is fixed at 10 time units.

\section{Spectroscopy}\label{Sec:Spec}

On the lattice, continuum SO(3) rotational symmetry is broken to the more restricted symmetry of the cubic group (also known as the octahedral group). Hadronic states at rest are classified according to irreducible representations (irreps) of the cubic group. Since cubic symmetry is respected by the lattice action, an operator belonging to a particular irrep will not mix with states in other irreps. The most general baryon spectrum from a lattice calculation is given by a small number of irreps of the double-cover of the cubic group: $G_{1 g/u}$, $H_{g/u}$ and $G_{2 g/u}$, where $g$ (German: gerade) and $u$ (ungerade) denote positive and negative parity, respectively.

We follow the technique introduced in Ref.~\cite{Basak:2005aq} to construct all the possible baryon interpolating operators that can be formed from local or quasi-local $u/d$ and $s$ quark fields; the $G_{2g/u}$ irreps are excluded, since they require non-(quasi-)local operators. In this paper, we present the calculations of the masses of the lowest-lying states in the $G_{1 g/u}$ and $H_{g/u}$ representations.

The non-locality in four dimensions of our valence DWF action manifests in oscillations of the two-point effective mass close to the source. As a result, a phenomenological form for fitting such data was proposed in Ref.~\cite{Syritsyn:2007mp}, employing both an oscillating contribution describing the non-local lattice artifacts and two positive-definite contributions:
\begin{eqnarray}
C(t) &=& A_0 e^{- M_0 (t - t_{\rm src})} + A_1 e^{- M_1 (t - t_{\rm
    src})}\nonumber\\ & +& A_{\rm osc} (-1)^t e^{- M_{\rm osc} (t - t_{\rm
    src})}.\label{eq:osc}
\end{eqnarray}
The first excited-state mass $M_1$ is included to extract a better ground-state mass $M_0$; $M_{\rm osc}$ is a non-physical oscillating term due to lattice artifacts. We use two-point correlators with the same smearing and interpolating operators at both source and sink and select fitted results with varying fit ranges, optimized for quality of fit. A standard jackknife analysis is employed here.

\begin{figure}
\includegraphics[width=0.45\textwidth]{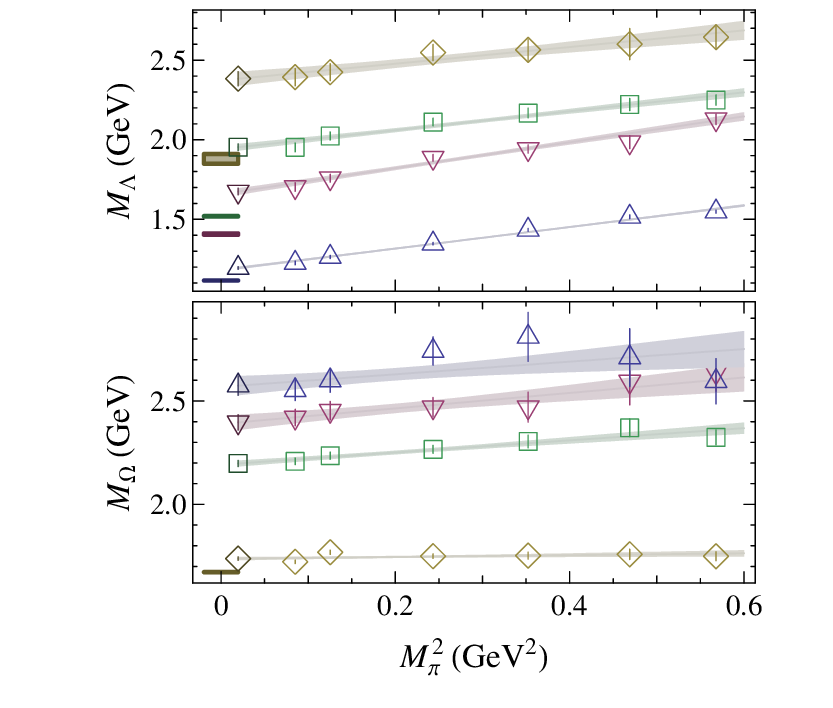}
\vspace{-0.3in}
\caption{The squared pion-mass dependence of $\Lambda$ and $\Omega$-flavor baryons with their extrapolated values. The bar at the left indicates the experimental values for the corresponding spin.
}\label{fig:spec-mpi2}
\end{figure}

Here we will focus on orbitally excited hyperon resonances. The ground states of the octet and decuplet and the SU(3) Gell-Mann--Okubo mass relation can be found in Ref.~\cite{WalkerLoud:2008bp}. There is in this reference an extensive description of baryon-mass extrapolation (to the physical pion mass) using continuum and mixed-action heavy-baryon chiral perturbation theory for two and three dynamical flavors. However, no chiral perturbation theory results have been published for orbitally excited hyperon resonances; therefore, we apply naive linear extrapolations in terms of $M_\pi^2$. Figure~\ref{fig:spec-mpi2} summarizes our results for the lowest-lying $\Lambda$ and $\Omega$ $G_{1g/u}$ (upward/downward-pointing triangles) and $H_{g/u}$ (diamonds and squares) (since these have less overlap with the data in in Ref.~\cite{WalkerLoud:2008bp}) and their chiral extrapolations. The leftmost points are extrapolated masses at the physical pion mass, and the horizontal bars are the experimental masses (if they are known).

\begin{figure}
\includegraphics[width=0.45\textwidth]{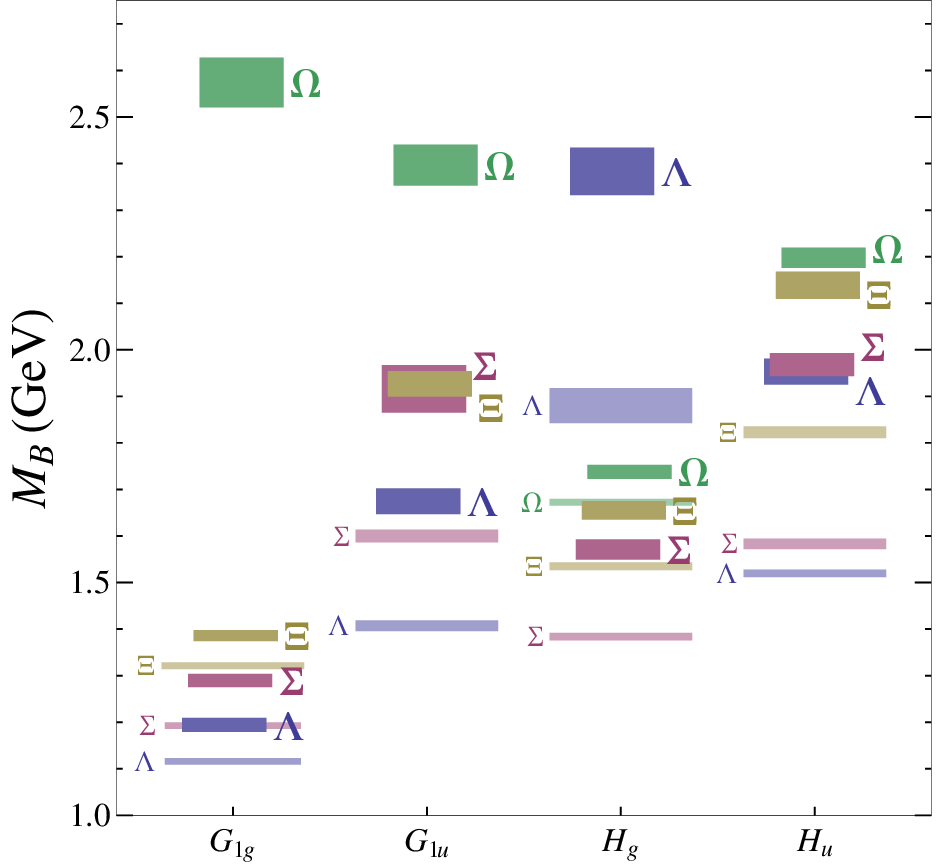}
\vspace{-0.3in}
\caption{A summary of our hyperon spectrum compared to experiment. Our data are the short bars with large labels on the right. The experimental states are long bars with small labels on the left.
}\label{fig:spec-barplot}
\end{figure}

We summarize the lattice hyperon-mass calculations for the $\Sigma$, $\Lambda$, $\Xi$ and $\Omega$ in Figure~\ref{fig:spec-barplot}, divided into vertical columns according to their discrete lattice spin-parity irreps ($G_{1g/u}$ and $H_{g/u}$), along with experimental results by subduction of continuum $J^P$ quantum numbers onto lattice irreps. $G_1$ ground states only overlap with spin-$1/2$. The spin identification for $H$ can be a bit trickier, since this irrep could match either to spin-$3/2$ or $5/2$ ground states. We simply select the lowest-lying of $3/2$ or $5/2$ indicated in the PDG, so it could be either depending on which one is the ground state for a particular baryon flavor. (We note below if the lowest $H$ is not spin-$3/2$.)

Although our naive extrapolation neglects contributions at next-to-leading-order chiral perturbation theory, several interesting patterns are seen.
The better-known $\Lambda$ (with $H_g$ being spin-$5/2$) and $\Sigma$ spectra match up with our calculations well. $G_{1u}$ $\Xi$ lines up well with the $\Xi(1690)$, indicating that the spin-parity of this resonance could be $1/2^-$; this agrees with SLAC's recent spin measurement.
The spin assignments for the $\Omega$ channel are the least known; from our mass pattern, we predict that $\Omega(2250)$ is likely to be $3/2^-$ (although we cannot rule out the possibility of $5/2^-$), and $\Omega(2380)$ and $\Omega(2470)$ are likely spin $1/2^-$ and $1/2^+$ respectively.

To successfully extract any reliable radially excited hyperon states, we would need finer lattice spacing, which would require massive computational resources to achieve. In Euclidean space the excited signals exponentially decay faster than the ground state. One possible solution is to use an anisotropic lattice, where the temporal lattice spacing is made finer than the spatial ones to reduce overall costs. A full-QCD calculation using anisotropic lattices is underway, and we would expect new results on these excited states within the next couple of years.

\section{Axial Coupling Constants}\label{Sec:AxialCouplings}

The hyperon axial couplings are important parameters entering the low-energy effective field theory description of the octet baryons. At the leading order of SU(3) heavy-baryon chiral perturbation theory, these coupling constants are linear combinations of the universal coupling constants $D$ and $F$, which enter the chiral expansion of every baryonic quantity, including masses and scattering lengths. These coupling constants are needed in the effective field theory description of both the non-leptonic decays of hyperons, and the hyperon-nucleon and hyperon-hyperon scattering phase shifts\cite{Beane:2003yx}. Hyperon-nucleon and hyperon-hyperon interactions are essential in understanding the physics of neutron stars, where hyperon and kaon production may soften the equation of state of dense hadronic matter.

\begin{figure}[t]
\begin{center}
\includegraphics[width=0.5\textwidth]{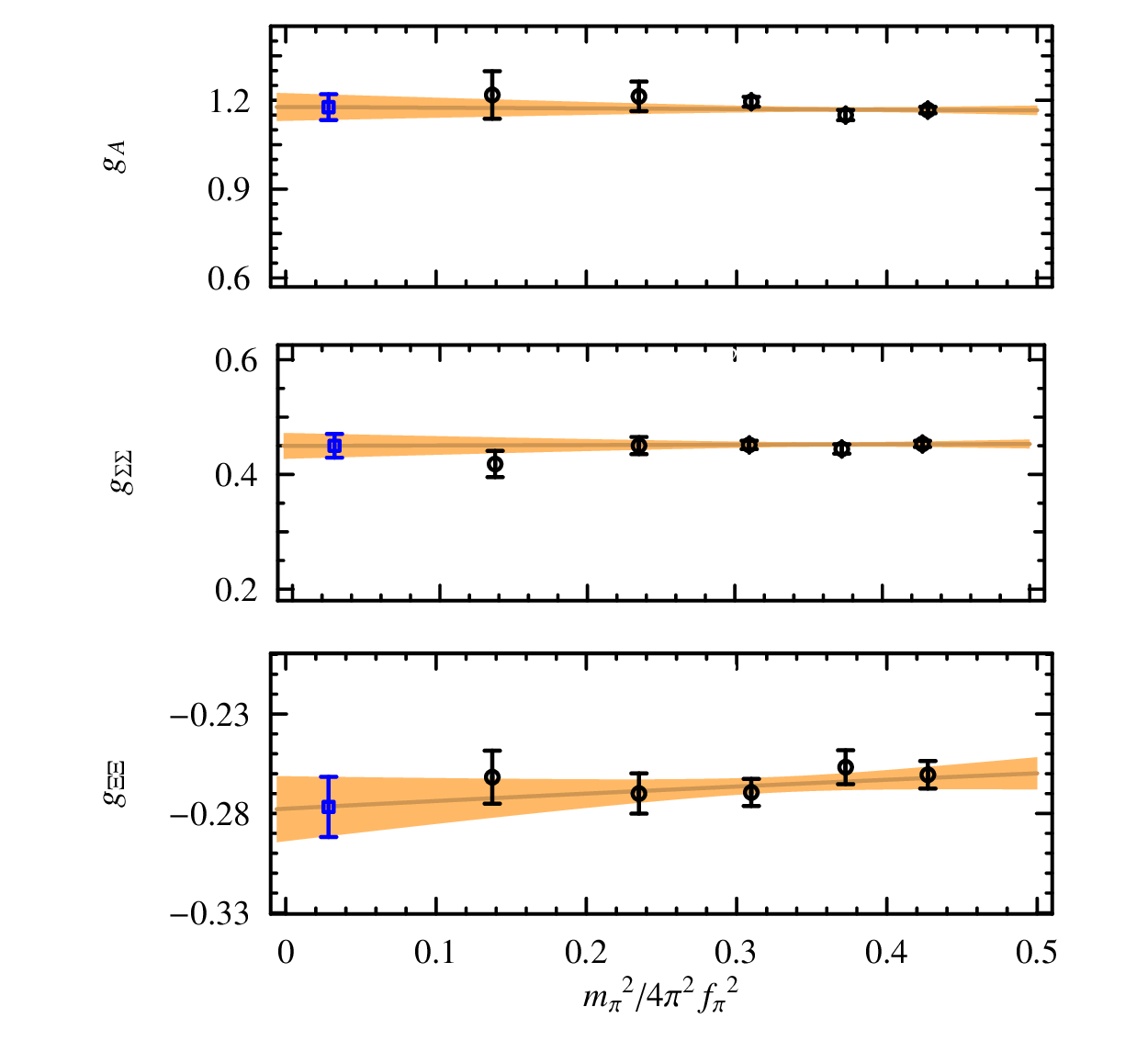}
\end{center}
\vspace{-0.3in}
\caption{Lattice data (circles) for $g_{A}$, $g_{\Sigma\Sigma}$ and $g_{\Xi\Xi}$ and chiral extrapolation (lines and bands). The square is the extrapolated value at the physical point.
}\label{fig:hyperonAxial}
\end{figure}

We have calculated the axial coupling constants for $\Sigma$ and $\Xi$ strange baryons using lattice QCD for the first time. We have done the calculation using 2+1-flavor staggered dynamical configurations with pion mass as light as 350~MeV. Figure~\ref{fig:hyperonAxial} shows our lattice data as a function of $(M_\pi/f_\pi)^2$ with the corresponding chiral extrapolation; the band shows the jackknife uncertainty. We conclude that $g_{A}=1.18(4)_{\rm stat}(6)_{\rm syst}$, $g_{\Sigma\Sigma}=0.450(21)_{\rm stat}(27)_{\rm syst}$ and $g_{\Xi\Xi} = -0.277(15)_{\rm stat}(19)_{\rm syst}$. In addition, the $SU(3)$ axial coupling constants are estimated to be $D=0.715(6)_{\rm stat}(29)_{\rm syst}$ and $F=0.453(5)_{\rm stat}(19)_{\rm syst}$. The axial charge couplings of $\Sigma$ and $\Xi$ baryons are predicted with significantly smaller errors than estimated in the past.

\begin{figure}
\begin{center}
\includegraphics[width=0.5\textwidth]{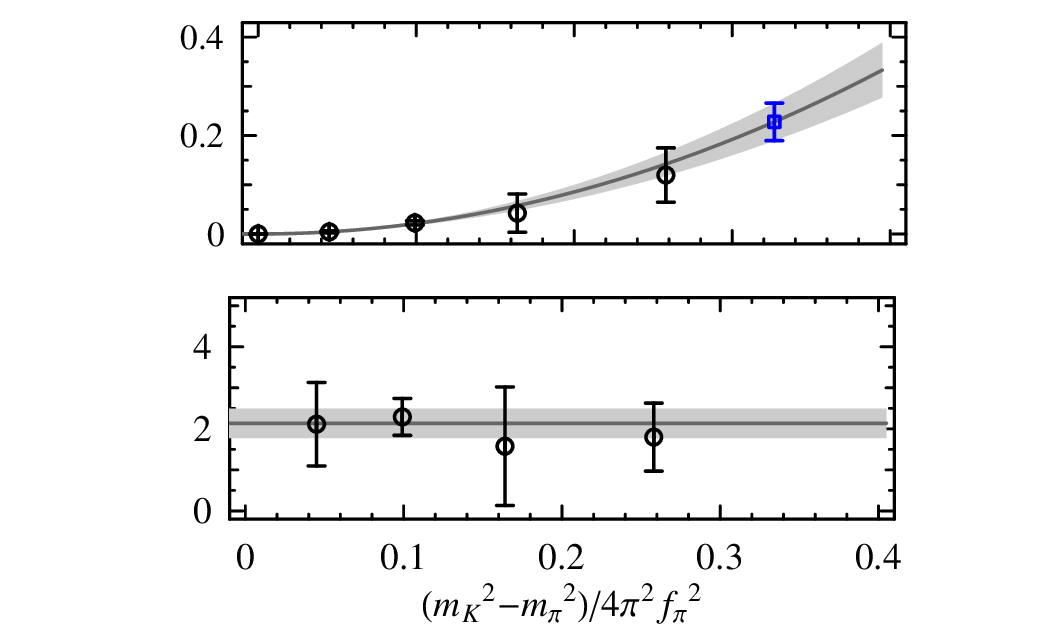}
\end{center}
\vspace{-0.3in}
\caption{(Top) The SU(3) symmetry breaking measure $\delta_{\rm SU(3)}$. The circles are the measured values at each pion mass, the square is the extrapolated value at the physical point, and the shaded region is the quadratic extrapolation and its error band. %\\
(Bottom) $\delta_{\rm SU(3)}/x^2$ plot. Symbols as above, but the band is a constant fit.}
\label{fig:SU(3)-breaking}
\end{figure}

We also study the SU(3) symmetry breaking in the axial couplings through the quantity $\delta_{\rm SU(3)}$,
\begin{equation}
\delta_{\rm SU(3)} = g_{\rm A}- 2.0\times g_{\Sigma\Sigma} +g_{\Xi\Xi} = \sum_n c_n x^n,
\label{eq:SU3break}
\end{equation}
where $x$ is ${(M_K^2-M_\pi^2)}/{(4\pi f_\pi^2)}$. Figure~\ref{fig:SU(3)-breaking} shows $\delta_{\rm SU(3)}$ as a function of $x$. Note that the value increases monotonically as we go to lighter pion masses. Our lattice data suggest that a $\delta_{\rm SU(3)} \sim x^2$ dependence is strongly preferred, as the plot of $\delta_{\rm SU(3)}/{x^2}$ versus $x$ in Figure~\ref{fig:SU(3)-breaking} also demonstrates. A quadratic extrapolation to the physical point gives 0.227(38), telling us that SU(3) breaking is roughly 20\% at the physical point, where $x = 0.332 $ using the PDG values\cite{PDBook} for $M_{\pi^+}$, $M_{K^+}$ and $f_{\pi^+}$. We compare the result of heavy-baryon SU(3) chiral perturbation theory\cite{Detmold:2005pt}  for $\delta_{\rm SU(3)}$ as a function of $x$, and we find that the coefficient of the linear term in Eq.~\ref{eq:SU3break} does not vanish. This implies that an accidental cancellation of the low-energy constants is responsible for this behavior.

\section{Form Factors}\label{Sec:FF}

The study of the hadron electromagnetic form factors reveals information important to our understanding of hadronic structure.
The electromagnetic form factors of an octet baryon $B$ can be written as
\begin{equation}\label{eq:lat_ME}
 \langle B\left|V_{\mu}\right|B\rangle
= {\overline u}_{B}
\left[\gamma_{\mu} F_1(q^2) +
\sigma_{\mu \nu}q_{\nu} \frac{F_2(q^2)}{2M_{B}}
\right]u_{B}\nonumber
\end{equation}
from Lorentz symmetry and vector-current conservation. $F_1$ and $F_2$ are the Dirac and Pauli form factors. Another common form-factor definition, widely used in experiments, are the Sachs form factors; these can be related to the Dirac and Pauli form factors through
\begin{eqnarray}\label{eq:lat-Gs}
G_E (q^2) &=& F_1(q^2) - \frac{q^2}{4M_B^2}F_2(q^2) \\
G_M (q^2) &=& F_1(q^2) + F_2(q^2).
\end{eqnarray}
In this work, we will concentrate on Sachs form factors. Note that here we only calculate the ``connected'' diagram, which means the inserted quark current is contracted with the valence quarks in the baryon interpolating fields.

On the lattice, we calculate the quark-component inserted current,  $V_{\mu}=\overline{q}\gamma_\mu q$, with $q=u,d$ for the light-quark current and $q=s$ for the strange-quark vector current. A single interpolating field for the nucleon, Sigma and cascade octet baryons has the general form
\begin{equation}\label{eq:lat_B-op}
 \chi^B (x) =  \epsilon^{abc} [q_1^{a\mathrm{T}}(x)C\gamma_5q_2^b(x)]q_1^c(x),
\end{equation}
where $C$ is the charge conjugation matrix, and $q_1$ and $q_2$ are any of the quarks $\{u,d,s\}$. For example, to create a proton, we want $q_1=u$ and $q_2=d$; for the $\Xi^{-}$, $q_1=s$ and $q_2=d$. By calculating two-point and three-point correlators on the lattice with the same baryon operator, we will be able to extract the form factors from Eq.~\ref{eq:lat-Gs}. We solve for $G_{M,E}$ using singular value decomposition (SVD) at each time slice from source to sink with data from all momenta with the same $q^2$ and all $\mu$.

%%%%%%%%%%%%%%%%%%%%%%%%%%%%%%%% G_E %%%%%%%%%%%%%%%%%%%%%%%%%%%%%%%%%

\begin{figure}
\includegraphics[width=0.45\textwidth]{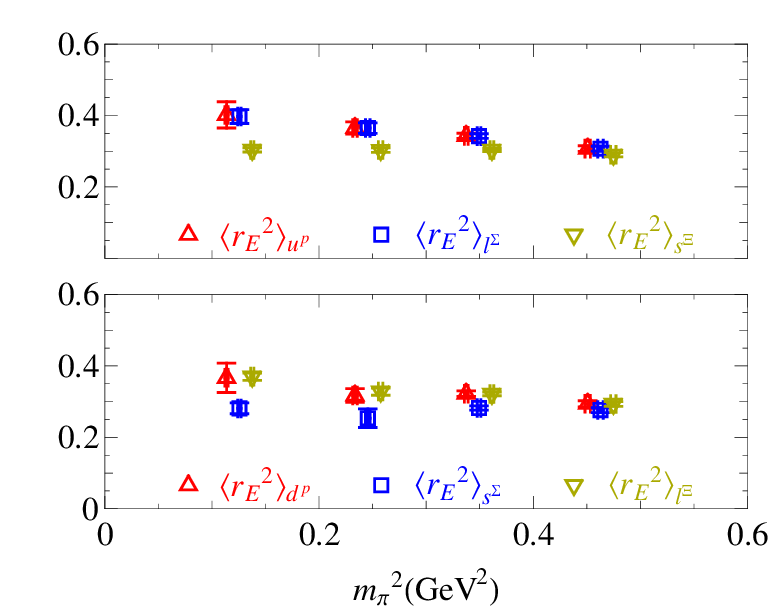}
\vspace{-0.3in}
\caption{The electric mean-squared radii in units of $\mbox{fm}^2$ as functions of $M_\pi^2$ (in GeV$^2$) from each quark contribution
}\label{fig:allB-r2E_mpi2}
\end{figure}

We constrain the fit form to go asymptotically to $1/Q^4$ at large $Q^2$ and to have $G_E(Q^2=0)=1$:
\begin{equation}
G_E = \frac{A Q^2+1}{C Q^2+1}\frac{1}{(1+Q^2/M_e^2)^2}.
\end{equation}
By trying various combinations of fit constraints on our data with squared momentum transfer  ${}<2$~GeV$^2$, we find $C$ is always consistent with 0; therefore, we set $C=0$. The mean-squared electric charge radii can be extracted from the electric form factor $G_E$ via
\begin{equation}\label{eq:GEradii}
\langle r_{E}^2\rangle = (-6)\frac{d}{dQ^2}\left(\frac{G_{E}(Q^2)}{G_{E}(0)}\right)\Big|_{Q^2=0}.
\end{equation}
In Figure~\ref{fig:allB-r2E_mpi2} we plot the electric charge radii with the neutron and $\Xi^0$ omitted, since the vector conserved current gives $G_{E,\{n,\Xi^0\}}(Q^2) \approx 0$ for these. We see that there is small SU(3) symmetry breaking between the SU(3) partners $p$ and $\Sigma^+$ (or $\Sigma^-$ and $\Xi^-$); their charge radii are consistent within statistical errors. Overall, the SU(3) symmetry breaking in the charge radii is much smaller than what we observed in our study of the axial coupling constants; for charge radii, the effect is negligible.

We can take a ratio of the electric radii of the baryons which coincide in the SU(3) limit; for example, $p$ and $\Sigma^+$ and $\Sigma^-$ and $\Xi^-$. The dominant meson-loop contribution is suppressed; thus, a naive linear fit to a ratio could be a better description than fitting individual channels. Following Sec.~\ref{Sec:AxialCouplings}, we use the SU(3) symmetry measure $x$ to parametrize the deviation of the ratio from 1 due to symmetry breaking:
$1+ \sum_{n=1}^N c_n x^n,$ where the next-order corrections contribute at the order of $x^{N+1}$; taking $n=1$, we expect the remaining effect should be less than 1\% in the expansion. The fit works fairly well for the extrapolation of $\frac{\langle r_E^2\rangle_{p}}{\langle r_E^2\rangle_{\Sigma^+}}$  and $\frac{\langle r_E^2\rangle_{\Sigma^-}}{\langle r_E^2\rangle_{\Xi^-}}$. By using the experimental value of $\langle r_E^2\rangle_{p}$ and $\langle r_E^2\rangle_{\Sigma^-}$, we can make predictions for $\langle r_E^2\rangle_{\Sigma^+}$ and $\langle r_E^2\rangle_{\Xi^-}$: $0.93(3)$ and $0.501(10)~\mbox{ fm}^2$ respectively. (Using $n=2$ in for the fit yields $c_2$ zero within error and thus gives results consistent with the extrapolation using $n=1$.)

%%%%%%%%%%%%%%%%%%%%%%%%%%%%%%%% G_M %%%%%%%%%%%%%%%%%%%%%%%%%%%%%%%%%
Studying the momentum-transfer dependence of magnetic form factors gives us the magnetic moment via
\begin{equation}\label{eq:magMoment}
\mu_B =  G_{M}^B(Q^2=0)
\end{equation}
with natural units $\frac{e}{2M_B}$, where $M_B$ are the baryon ($B\in\{N,\Sigma,\Xi\}$) masses. To compare among different baryons, we convert these natural units into nuclear magneton units $\mu_N=\frac{e}{2M_N}$; therefore, we convert the magnetic moments with factors of $\frac{M_N}{M_B}$.

\begin{figure}
\includegraphics[width=0.45\textwidth]{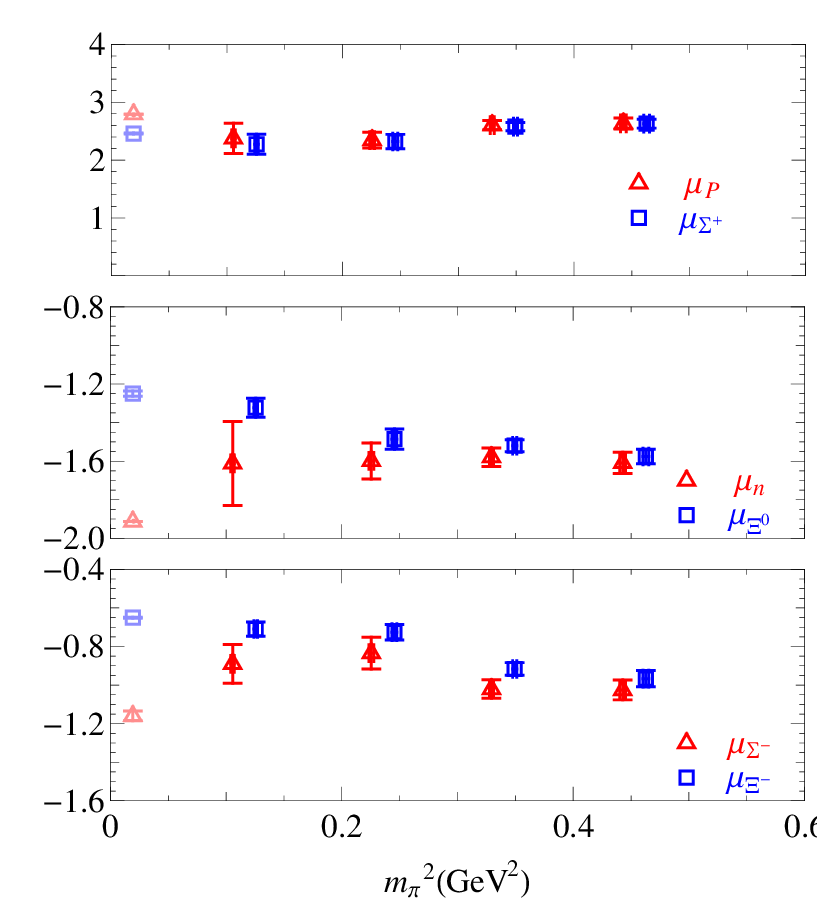}
\vspace{-0.3in}
\caption{Baryon magnetic moments in units of $\mu_N$ as functions of $M_\pi^2$ (in GeV$^2$).  The leftmost points are the experimental numbers.
}\label{fig:allB-muMrErM_mpi2}
\end{figure}

We can obtain the magnetic moments and radii from polynomial fitting to the ratio of magnetic and electric form factors, $G_M/G_E$. From the definition of the electric and magnetic radii, we expect that $G_M/G_E \approx A +C Q^2$, where the magnetic moment is $\mu=A G_E(0)$, and $C$ is proportional to $\langle r_M^2 \rangle-\langle r_E^2 \rangle$. In the case of $n$ and $\Xi^0$, we use $G_{E,p}$ and $G_{E,\Xi^-}$ in the ratio instead of $G_{E,n}$ and $G_{E,\Xi^0}$. Figure~\ref{fig:allB-muMrErM_mpi2} shows the magnetic moments of each baryon compared with its SU(3) partner: $\{p,\Sigma^+\}$, $\{n,\Xi^0\}$, $\{\Sigma^-,\Xi^-\}$. We find that as seen in experiment, the SU(3) breakings of the magnetic moments are rather small. As we go to larger pion masses (that is, as the light mass goes to the strange mass), the discrepancy gradually goes to zero as SU(3) is restored. But even at our lightest pion mass, around~350 MeV, the effects of SU(3) symmetry breaking effect can be ignored. The fitted results are consistent with what we obtained from the dipole extrapolations. We examine the radii differences from the quark contributions and observe less than 10\% discrepancy.
The ratio approach also benefits from cancellation of noise due to the gauge fields, and thus it has smaller statistical error. Therefore, we will concentrate on the results from this approach for the rest of this work.

We extrapolate the baryon magnetic moments using the SU(3) ratios with the SU(3) symmetry breaking measure $x$. Again, the ratio has cleaner statistical signal due to the cancellation of fluctuations within gauge configurations, and the linear extrapolation is not a bad approximation, since potential log terms are suppressed. With the help of experimental $\mu_{\Sigma^+}$, $\mu_{\Xi^-}$ and $\mu_{\Xi^0}$\cite{PDBook}, we obtain $\mu_p=2.56(7)$, $\mu_n=-1.55(8)$ and $\mu_{\Sigma^-}=-1.00(3)$. (The fit using up to $x^2$ terms results in a zero-consistent fit parameter $c_2$ and yields numbers consistent with the above.)

SU(6) symmetry predicts the ratio $\mu_{d^p}/\mu_{u^p}$ should be around $-1/2$. Compared with what we obtain in this work, the ratio agrees within $2\sigma$ for all the pion-mass points. The heaviest two pion points have roughly the same magnitude as in the quenched calculation\cite{Boinepalli:2006xd}. However, at the lightest two pion masses, they are consistent with the $-1/2$ value. The difference could be due to sea-quark effects, which become larger as the pion mass becomes smaller. A naive linear extrapolation through all the points gives $-0.50(10)$. SU(6) symmetry is preserved in the lattice calculations.

We also check the sum of the magnetic moments of the proton and neutron, $\mu_{p}+\mu_{n}$, which should be about 1 from isospin symmetry. Again, the values from different pion masses are consistent with each other within 2 standard deviations and differ from 1 by about the same amount. A naive linear extrapolation suggests the sum is 0.78(13), which is consistent with experiment but about $2\sigma$ away from 1. This symmetry is softly broken, possibly due to finite lattice-spacing effects. Finer lattice-spacing calculations would be needed to confirm this. 

\section{Semileptonic Decays}\label{Sec:Semi}

The hyperons differ from the nucleons by their strangeness, and although hyperon decays via the weak interaction have been known for more than half a century, interest in their study has not decayed with time. They provide an ideal systems in which to study SU(3) flavor symmetry breaking and offer unique opportunities to understand baryon structure and decay mechanisms. The low-energy contribution to the transition matrix elements for hyperon beta decay, $B_1 \rightarrow B_2 e^- \overline{\nu}$ can be written in general form as
\begin{eqnarray}\label{eq:ME}
{\cal M} =\frac{G_s}{\sqrt{2}}\overline{u}_{B_2}(O_\alpha^{\rm
V}+O_\alpha^{\rm
A}){u}_{B_1}\overline{u}_{e}\gamma^\alpha(1+\gamma_5)v_{\nu}.
\end{eqnarray}
From Lorentz symmetry, we expect the matrix element composed of any two spin-$1/2$ nucleon states, $B_1$ and $B_2$, to have the form
\begin{equation}\label{eq:cont_O's}
O_\alpha^{V} = f_1(q^2)\gamma^\alpha
                 + \frac{f_2(q^2)}{M_{B_1}}\sigma_{\alpha\beta}q^\beta
                 + \frac{f_3(q^2)}{M_{B_1}}q_\alpha \nonumber%\\
\end{equation}
\begin{equation}
O_\alpha^{A} = \left(g_1(q^2) \gamma^\alpha
                 + \frac{g_2(q^2)}{M_{B_1}}\sigma_{\alpha\beta}q^\beta
                 + \frac{g_3(q^2)}{M_{B_1}}q_\alpha\right)\gamma_5  \nonumber
\end{equation}
with transfer momentum $q=p_{B_2}-p_{B_1}$ and $V,A$ indicating the vector and axial currents respectively. $f_1$ and $g_1$ are the vector and axial form factors, and $f_2$ and $g_2$ are the weak magnetic and induced-pseudoscalar form factors. They are non-zero even with $B_1=B_2$. SU(3) flavor symmetry breaking accounts for the non-vanishing induced scalar and weak electric form factors, $f_3$ and $g_2$ respectively. Due to the difficulty in disentangling experimental form factor contributions, they tend to be set to zero. We will be able to determine these form factors using theoretical technique and will find them to be non-negligible.

So far, there are only two ``quenched'' (where the fermion masses in the sea sector are infinitely heavy) lattice calculations of hyperon beta decay, and they are in different channels, $\Sigma \rightarrow n$ and $\Xi^0 \rightarrow \Sigma^+$.
Guadagnoli~et~al.\cite{Guadagnoli:2006gj} calculated the matrix element $\Sigma \rightarrow n$ with all of the pion masses larger than 700~MeV. Sasaki~et~al.\cite{Sasaki:2006jp} used lighter pion masses in the range 530--650~MeV and DWF to look at the $\Xi^0$ decay channel. They extrapolate the vector form factor $f_1$ using the parameter $\delta=(M_{B_2}-M_{B_1})/M_{B_2}$. In this work, we calculate both decay channels and remove the quenched approximation, which often causes notoriously large systematic error. Our fermion sea sector contains degenerate up and down quarks plus the strange. We use pion masses as light as 350~MeV, which ameliorates some of the uncertainty in the extrapolation to the physical pion mass. To condense our work for this proceeding, we will concentrate on the results from $\Sigma \rightarrow n$.

\begin{figure}
\vspace{-0.3in}
\includegraphics[width=0.46\textwidth]{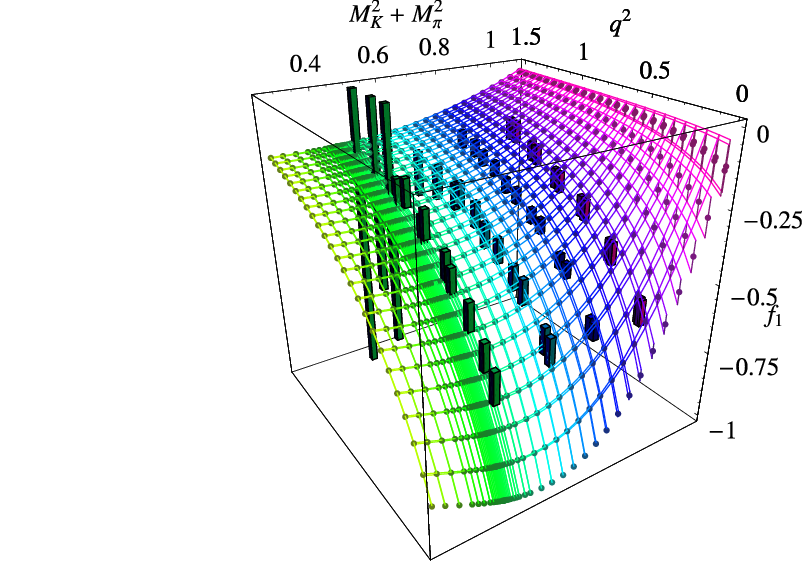}
\vspace{-0.3in}
\caption{Simultaneous extrapolation in $q^2$ and mass }\label{fig:f2dmq}
\end{figure}

To obtain $V_{us}$, we need to extrapolate $f_1$ to zero momentum-transfer (we cannot calculate this point directly due to the discrete values of momentum accessible in a finite volume) and the physical pion and kaon masses. Fortunately, $f_1$ is protected by the Ademollo-Gatto (AG) theorem such that there is no first-order SU(3) breaking. Therefore, the quantity deviates from its SU(3) value by the order of the symmetry breaking term of $O({H^\prime}^2)$, where $H^\prime$ is the SU(3) symmetry breaking Hamiltonian; the natural candidate an observable to track this breaking is the mass splitting between the kaon and pion. Combining with momentum extrapolation (using a dipole form in this case), we use a single simultaneous fit:
\begin{equation}\label{eq:CombinedFit}
f_1(q^2) = \frac{1+\left(M_K^2-M_\pi^2\right)^2
\left(A_1+A_2\left(M_K^2+M_\pi^2\right)\right)}
{\left(1-\frac{q^2}{M_0+M_1\left(M_K^2+M_\pi^2\right)}\right)^2}.
\end{equation}
Figure~\ref{fig:f2dmq} shows the result from simultaneously fitting over all $q^2$ and mass combinations for the $\Sigma^-\rightarrow n$ decay. The $z$-direction indicates $f_1$, while the $x$- and $y$-axes indicate mass and transfer momentum. The surface is the fit using Eq.~\ref{eq:CombinedFit} with color to indicate different masses. The columns are the data and the momentum points from different pion masses line up in bands. Our preliminary result for $f_1$ is $-0.95(3)$ (which is consistent with the quenched result\cite{Guadagnoli:2006gj}: $f_1=-0.988(29)_{\rm stat}$.)

%%%%%%%%%%%%%%%%%%%%%%%%%%%%%%%%%%%%%%%%%%%%%%%%%%%%%%%%%%%%%%%%%%%%%%%

The axial form factor $g_1$ is not protected by the AG theorem. Experimentally, one is interested in its ratio with the vector form factor, $g_1/f_1$, at zero momentum transfer. Here we adopt naive linear mass combinations $(M_K^2+M_\pi^2)$ and $(M_K^2-M_\pi^2)$ and momentum dependence to extrapolate and find $g_1(0)/f_1(0)=-0.336(52)$, which is consistent with the experimental value of $-0.340(17)$. %\cite{Cabibbo:2003cu}.
Similar extrapolations are applied to other form factor ratios, such as $f_2(0)/f_1(0)=-1.28(19)$, which is consistent with Cabibbo model value of $-1.297$\cite{Cabibbo:2003cu}.

\begin{figure}[h]
\includegraphics[width=0.45\textwidth]{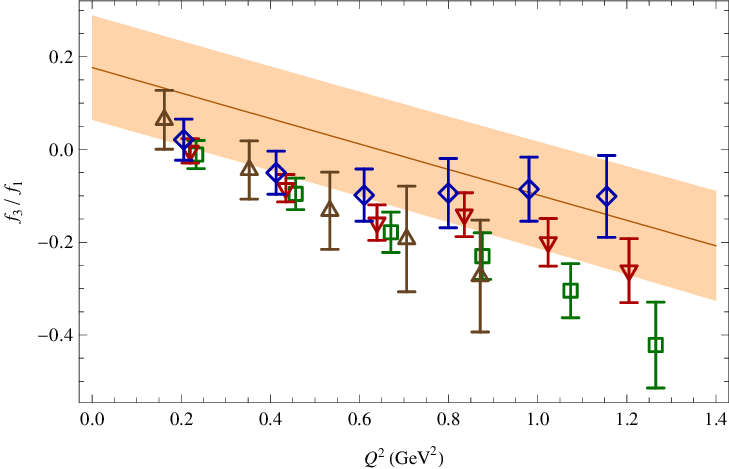}
\includegraphics[width=0.45\textwidth]{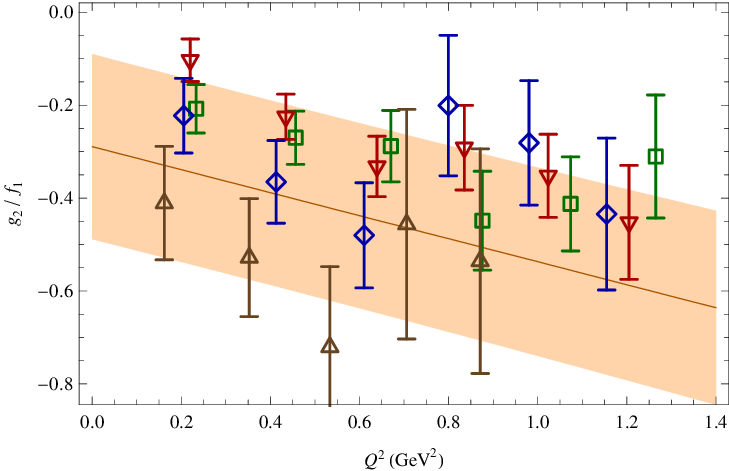}
\vspace{-0.3in}
\caption{The momentum dependence of the form factor ratios $f_3/f_1$ (top) and $g_2/f_1$ (bottom) with various pion masses labeled by triangles, squares, downward triangles and diamonds from light to heavy; the band indicates the extrapolation at the physical mass
}\label{fig:f3g2}
\end{figure}

Finally, we turn our discussion toward the SU(3)-vanishing weak-electric $g_2$ and induced-scalar $f_3$ form factors. Figure~\ref{fig:f3g2} presents the momentum dependence of the ratios $g_2(q^2)/f_1(q^2)$ and $f_3(q^2)/f_1(q^2)$ for sea-pion masses ranging 350--700~MeV. The band indicates our mass extrapolation in terms of a naive linear dependence of $M_K^2-M_\pi^2$ and $M_K^2+M_\pi^2$. We find $f_3(0)/f_1(0)=-0.17(11)$ and $g_2(0)/f_1(0)=-0.29(20)$, which are 1.5 standard deviations from zero. Experimentally, only a combination of axial form factors $\left|g_1(0)/f_1(0)-0.133 g_2(0)/f_1(0)\right|$ is determined. They find 0.327(7)(19), which is consistent with our result, $0.297(60)$.

\section{Summary and Outlook}\label{Sec:End}

The lattice formulation is a powerful method for calculating nonperturbative quantities used in fundamental tests of QCD. The results presented here demonstrate clear progress toward the long-term goals of determining the hyperon spectrum, structure and decays from first principles. Ongoing efforts from lattice community using improved operators and algorithms (to improve the noise-to-signal ratios), finer lattice spacings (to reduce systematic error due to discretization), lighter pion masses (to reduce uncertainty introduced by chiral extrapolation) on dynamical sea quarks will continue to better approximate the physical world and provide observables that elude experiments.

\section*{Acknowledgements}
HWL thanks collaborators Kostas Orginos, David Richards, Colin Morningstar and Saul Cohen for useful discussions.  These calculations were performed using the Chroma software suite\cite{Edwards:2004sx} on clusters at Jefferson Laboratory using time awarded under the SciDAC Initiative. Authored by Jefferson Science Associates, LLC under U.S. DOE Contract No. DE-AC05-06OR23177. The U.S. Government retains a non-exclusive, paid-up, irrevocable, world-wide license to publish or reproduce this manuscript for U.S. Government purposes.

%\bibliographystyle{JHEP}
%\bibliography{nuc_ref}
\providecommand{\href}[2]{#2}\begingroup\raggedright\endgroup

\end{document}